\def\rf#1;#2;#3;#4;#5 {\par#1, {\it #3} {\bf #4}, #5 (#2). \par}
\def\beq#1{\begin{equation}\label{#1}}
\def\eeq{\end{equation}}
\def\beqa#1{\begin{eqnarray}\label{#1}}
\def\eeqa{\end{eqnarray}}

\def\ignore#1{}
\def\refe{\par\noindent\hangindent=1.5cm}

\def\spose#1{\hbox to 0pt{#1\hss}}
\def\simlt{\mathrel{\spose{\lower 3pt\hbox{$\mathchar"218$}}
     \raise 2.0pt\hbox{$\mathchar"13C$}}}
\def\simgt{\mathrel{\spose{\lower 3pt\hbox{$\mathchar"218$}}
     \raise 2.0pt\hbox{$\mathchar"13E$}}}
\def\simpropto{\mathrel{\spose{\lower 3pt\hbox{$\mathchar"218$}}
     \raise 2.0pt\hbox{$\propto$}}}

\def\ed{\end{document}}
\def\bl{\vskip0.423truecm}
\def\ind{\noindent\hskip1.8truecm}

\documentclass [11pt]{article}

\begin{document}
\def\basel{\normalsize\baselineskip=0.85truecm}
{\hfill} \vskip0.7truecm

\noindent {\bf IS QUANTUM SUICIDE PAINLESS? ON AN APPARENT
VIOLATION OF THE PRINCIPAL PRINCIPLE}

\bl\bl\bl \ind Milan M. \'Cirkovi\'c

\bl\ind {\it Astronomical Observatory Belgrade

}

\ind {\it Volgina 7, 11160 Belgrade, Serbia and Montenegro}

%\ind {\it Serbia and Montenegro}

\ind {\tt mcirkovic@aob.bg.ac.yu}

\bl\bl\bl\ind Received ...; accepted ...

\bl\bl

\baselinestretch

\noindent {\bf Abstract.} The experimental setup of the
self-referential quantum measurement, jovially known as the
"quantum suicide" or the "quantum Russian roulette" is analyzed
from the point of view of the Principal Principle of David Lewis.
It is shown that the apparent violation of this
principle---relating objective probabilities and subjective
chance---in this type of thought experiment is just an illusion
due to the usage of some terms and concepts ill-defined in the
quantum context. We conclude that even in the case that Everett's
(or some other "no-collapse") theory is a correct description of
reality, we can coherently believe in equating subjective credence
with objective chance in quantum-mechanical experiments. This is
in agreement with results of the research on personal identity in
the quantum context by Parfit and Tappenden.

%\vspace{1cm}
\section{Introduction: Principal Principle in the multiverse}

\noindent The so-called Principal Principle (henceforth PP)
introduced by David Lewis (1986) suggests, roughly, that our
credence in a proposition A, given all admissible evidence, should
equate with one's best estimate of the objective chance that A
will come true. Precisely, Cr(A, XE) = x, where A is a
proposition, X is the proposition that the chance of A being true
is x, x is a real number between 0 and 1, E is any ``admissible''
evidence, and Cr(.) is a reasonable credence function. From this
general formulation we conclude that if Cr(XE) = 1 then Cr(A) = x.
Applied to physical experiments, PP tells us that we should
believe one of the possible outcomes of an experiment, say
A$_{i}$, with credence
\begin{equation}
\label{jjedan}
 {\rm Cr}({\rm A}_{i}) = p_{i},
\end{equation}
\noindent where p$_{i}$ is the physical chance of this outcome,
provided we possess only admissible information about the
experiment. This is especially convenient for experiments in
quantum mechanics, like the stock example of radioactive decay of
a heavy nucleus, since the underlying physical mechanism is
usually thought of as being irreducibly stochastic; thus, the
possibility of some inadmissible information slipping in unnoticed
is remote. (Refinements of the PP, criteria of admissibility, and
suggested alternative chance-credence relations often appear in
the literature, but we shall not consider them here.) Usually, PP
is just an embodiment of our common sense intuitions; we simply
follow (1) in everyday life as well as in physics, with routine
success. But, PP can be---at least apparently---violated if the
so-called ``no-collapse'' versions of quantum mechanics (e.g.\
Everett 1957; Deutsch 1985; Tegmark 1998) turn out to be correct.
Here is the recipe.

Let Hugh, an experimental physicist, prepare a simple coherent
state of, say, spin $z$-projection of a fermion:
\begin{equation}
\label{eq1} \left|\psi \right\rangle = \frac{1}{\sqrt 2 }\left(
{\left| \uparrow \right\rangle + \left| \downarrow \right\rangle }
\right),
\end{equation}
(any other quantum superposition will do equally well). The spin
measuring device is coupled to a revolver in such way that, after
Hugh or anybody else pulls the trigger, spin measurement will take
place. The measured eigenstate $\left| \downarrow \right\rangle $
will result in firing the revolver, and the measured $\left|
\uparrow \right\rangle $ will result in a harmless ``click''. One
important proviso is that the duration of measurement (including
the reaction time of the revolver) should be short in comparison
to any timescale characterizing human perception. If the revolver
is aimed at any target other than Hugh himself, he expects to see
a seemingly random outcome of each individual measurement: either
``bang'' or ``click'' with the probability of a coin toss. If we
suppose that measurements are taken in series of $n$ consecutive
measurements (pullings of the trigger), probability of achieving
any individual combination of ``bangs'' and ``clicks'' is given by
the simplest binomial distribution; in particular, the probability
of obtaining "click" $n$ consecutive times is $p(n \! \uparrow) =
0.5^n$. There is no observed---or observable---difference between
collapse and ``no-collapse'' quantum theories at this
point.\footnote{Metaphysically, Hugh is aware that the complete
description is certainly different in these two cases; notably, on
the ``no-collapse'' quantum mechanics he believes in, both a
``click'' and a ``bang'' will be realized in each measurement,
being the two components of the unbreakable superposition, but he
will perceive only one of them due to the rapid environmental
decoherence. The decoherence timescale is, even for quite isolated
fermion spins, still much shorter than the human perception
timescale.}

Things take a strange turn when Hugh decides---perhaps upon advice
of some respectable quantum theorists---to point the revolver to
his own head.\footnote{Hence the name of ``quantum Russian
roulette'' or ``quantum suicide''.} The state of the entire
system, after the first measurement, now evolved from (\ref{eq1})
to (symbolically)
\begin{eqnarray}
\label{ddva} \hat{U} \left| \psi \right\rangle \otimes \left| {\rm
experimenter} \right\rangle & = & \hat{U}\frac{1}{\sqrt{2}}
(\left| \uparrow \right\rangle+ \left| \downarrow
\right\rangle)\otimes \left| {\rm experimenter} \right\rangle = \\
& = & \frac{1}{\sqrt{2}} (\left| \uparrow \right\rangle \otimes
\left| {\rm "click"} \right\rangle + \left|\downarrow
\right\rangle \otimes \left| {\rm "dead"} \right\rangle).
\end{eqnarray}
It is almost self-evident that no probability (objective or
otherwise) makes sense for Hugh in his $\left| {\rm "dead"} \right
\rangle$ state (but see Papineau 2003 and comments below). Since
we have exactly one observer before and after the measurement, and
since the decoherence of branches---the two terms in parenthesis
in (3)---occurred much faster than our experimenter could notice,
we may be certain that he will observe spin "up", and
\textit{therefore\/} hear a harmless "click". And the same could
be repeated arbitrarily often! Notice that only physical
collapse---as in the "orthodox" Copenhagen interpretation or the
dynamical reduction theories---is actually harmful from Hugh's
point of view. Since in the "no-collapse" view there is no actual
collapse, just fast decoherence between the branches, Hugh will
find himself in the strange situation of impossibility of
committing suicide, although the revolver is loaded and fully
functional! This is different from the "outsider" view of, say,
assistant in the experiment, who will perceive the bloody deed
after at most several repetitions of the experiment (being in one
of the decohered branches of the universal wavefunction
\textit{and\/} being able to perceive the measured spin ``down'').

This is, of course, the ``quantum suicide'' experiment of Squires
(1986), Moravec (1988), Zeh (1992), Price (1996) and Tegmark
(1998)---it seems that many people have arrived independently
atthe same idea, although the late Euan Squires seems to be the
first who wrote about it,\footnote{Although a SF story of John
Gribbin preceded it for about a year: "The Doomsday Device",
published in Analog, January 22, 1985 (personal communications
with H. Moravec and J. Gribbin).} and only with Tegmark's (1998)
paper it received proper attention. Recently, it has attracted a
lot of attention in philosophical circles either (Lewis 2000,
2004; Papineau 2003, 2004; Tappenden 2004). Of course, different
variations of the experimental setup are possible. It can be even
cast in Moravec's form of the ``collective'' quantum suicide
(quantum genocide?), i.e.\ running a potentially catastrophic
experiment on the new super-duper accelerator. In all these cases
one thing is clear: while the objective chance of measuring spin
"up" in Hugh's single measurement is, of course, p$_{\uparrow}$ =
0.5, subjective probability of the same ("surviving") outcome is
Cr($ \uparrow )$ = 1. Obviously, this violates (1) and Hugh finds
PP misleading. Moreover, by repeating the measurement, our
experimenter can obtain arbitrary large PP violation (and,
simultaneously, obtain an arbitrarily large degree of confidence
in the validity of the no-collapse theory).

One rather obvious way out of the difficulty is to contest the
usual death-defying conclusion of the quantum suicide thought
experiment. This has been recently argued {\it in extenso}, though
with somewhat different motivation, by Papineau (2003, 2004). In
reply to a paper by Peter J.\ Lewis (2000), he proposes
effectively assigning a (negative) utility value to Hugh's $\left|
{\rm "dead"} \right \rangle$ state. Without going into details of
the argument (soon diverted to the issue of life-staking degree of
belief in physical theories which, though of paramount general
importance, is irrelevant for our purposes here), we notice at
least its rather {\it ad hoc\/} nature. Further, it has been, in
our view, strongly and correctly criticized by Tappenden (2004)
whose arguments, in conjunction with the original Lewis (2000)
claim, vindicate the legitimacy of the conventional quantum
suicide on most versions of Everettian theories. Of course, Peter
J.\ Lewis and Tappenden have quite different views on whether this
bodes well or not for such theories, but it is also less relevant
from our point of view. It would, of course, be more important and
interesting to discuss any issues related to the quantum
multiverse if a no-collapse theory turns out to be true, perhaps
by some radical new experiment like the one proposed by Deutsch
(1985); but even if things turn otherwise, the general puzzle
seems interesting enough.

\section{Analysis}

Should we reject PP because it seems apparently violated in the
quantum suicide experiment? Not really. After all, as with all
principles, PP should be understood primarily as a "guideline" in
our probabilistic considerations. It perceives objective chances
as concise, informative summaries of patterns of local empirical
facts. The violation described above arises because conventional
guidelines are unavoidably geared toward experiments performed
objectively, without real participation of an individual observer.
Subjectively defined outcomes are, naturally, somewhat more
difficult to analyze in the physical context; we shall show below
how we ought to actually analyze various positions in respect to
the issue of personal identity in the no-collapse theories, in
order to explain how PP remains valid. This seems like another
instance of the necessity of taking into account the ``anthropic
bias'' (cf.\ Bostrom 2002), necessarily limiting the entire space
of \textit{prima facie }possible experimental outcomes. Whenever
it is possible to infer the outcome of the chance experiment with
certainty, the value of objective chance becomes irrelevant. What
is so striking here is that ``no-collapse'' quantum theories give
us a perfectly scientific opportunity to reach such knowledge at
the fundamental level, without recourse to proverbial ``crystal
balls'' and other non-physical devices used in the literature to
exemplify inadmissible evidence (e.g.\ Vranas 1998). The simple
insight which the experimenter (if knowledgeable about the
``no-collapse'' theories, as Hugh certainly is!) possesses about
the outcome of the experiment entails that he will not believe the
PP-favored value of 50{\%} chance.

Let us consider now what exactly the controversial proposition is
in the above-described experiment. Obviously, in conjecturing that
PP is violated, our proposition has been the following:

\vspace{0.5cm}

A: \textit{Hugh will survive the quantum suicide experiment.}

\vspace{0.5cm}

Seemingly, this proposition violates PP, as stated above. However,
this is an illusion, stemming from uncritical use of our
conventional notions of existence and survival in the
inappropriate context of the no-collapse theories. Spin
measurement causes the universal wavefunction to ``branch''; what
happens to Hugh during the branching process? The answer to this
question depends on our theories of personal identity in the
multiverse, the issue recently clearly investigated by Tappenden
(2000; see also Parfit 1984). We have stated the necessity of
timescales for the apparatus and the gun to be restricted;
however, we can always find a small temporal interval between the
spin measurement and the actual firing/clicking of the gun, and
pose our question in that time interval. There are several at
least superficially plausible answers, notably: (i) Hugh exists in
all branches, (ii) Hugh exists in none of the branches, and (iii)
Hugh exists in some of the branches, but not in others. Let us
analyze these cases from the point of view of chances in the
quantum suicide experiment, and \textit{neglecting other
problematic features\/} some of these options may have for other
reasons.\footnote{For instance, the difficulties can consist in
the implied account of the tensed discourse. If Hugh survives in
one wavefunction branch and not in the other (case i.), doesn't
that make it the case that "Hugh will survive the quantum suicide
experiment" and "Hugh will not survive the quantum suicide
experiment" are both true, while their conjunction "Hugh will
survive the quantum suicide experiment and he will not survive the
quantum suicide experiment" is false?}

If Hugh exists in all branches, then we may be certain that he will survive
the experiment, and it is with this case in mind that Squires and others
have actually suggested the quantum suicide thought experiment in the first
place. This is in accordance with the idea of reductionism of Parfit or
person-reductionism of Tappenden, i.e.

\begin{quotation}
\noindent the idea that there is no more to the identity of a
person over time than that there is a series of personal states
linked by some relation, where personal states are identified in a
way which does not require their attribution to a person. A
personal state can be thought of as a group of mental
characteristics which all belong to what we would normally think
of as history of a single waking human body over a short period of
time. The duration of this period is the amount of time we think
of as being the present moment, the minimal time-frame in which we
have experiences; it is often called the specious present. ... We
can think of the duration of a typical personal state as being
something around a fifth of a second.\footnote{Tappenden (2000),
pp. 106-107.}
\end{quotation}

What sense do we make of the proposition A here? It is reasonable
to understand it as a claim that Hugh survives in \textit{at least
one world\/} (of the multiverse), which has a chance of 100{\%},
and as PP states, and Squires, Tegmark and other infer, Hugh
should be certain that the proposition is true. This, of course,
does not mean that Hugh should think that the fermion spin could
not be in the eigenstate $\left| \downarrow \right\rangle$, but
this is another proposition, different from A; we shall consider
this fine point below.

If the case (ii) is correct, then Hugh is substituted for some
other individual(s), say Hugh$_{1}$ in the branch in which the
spin is "up", and Hugh$_{2}$ in the branch where the spin is
"down". (This is, parenthetically, a nice illustration of the
Tappenden's concept of distinction along the \textit{superslice\/}
dimension. On the other hand, this case is antithetical to the
entire idea of the quantum suicide experiment, and is considered
here just for the sake of completeness.) However, now the
proposition A above is obviously wrong, since Hugh $\ne$
Hugh$_{1}$ and Hugh $\ne$ Hugh$_{2}$. Thus the chance of
proposition A being true is zero and, as PP states, Hugh should be
certain that he would not survive.

Finally, we have the case (iii), where we suppose that Hugh exists
in the wavefunction branch in which the gun clicks. In the other
one, presumably, he is replaced with (dead) individual Hugh$_{1}$.
Now, the chance of the proposition A is just the
quantum-mechanical probability 50{\%}, and, as PP says, Hugh
should have credence 50{\%} in this proposition. Note that this
case is completely analogous to the case of quantum suicide in the
collapse quantum theories.

We conclude that PP remains entirely valid. Now, why is this so
counter-intuitive? The answer is that we have confused the proposition A
above with the different proposition, say B, stating that

\vspace{0.5cm}

B: \textit{The result of spin measurement is the eigenstate
}$\left| \uparrow \right\rangle $.

\vspace{0.5cm}

In contrast to A, the proposition B obviously does not depend on
the issue of personal identity; it is completely objective
statement about physical system. However, it also can be
understood in various ways, depending on the sort of quantum
theory we accept. In the no-collapse theories considered here, we
may understand B to mean that the result of measurement will be
spin ``up'' in at least one world. This obviously has the chance
of 100{\%}, and agrees well with the outcome of our analysis of
the case (i) above for the proposition A. Now, if we define the
entire quantum suicide experimental setup as B $ \Rightarrow $ A,
there is no contradiction. Problems may arise when we interpret B
as the statement that the spin is ``up'' in \textit{this
particular branch of the universal wavefunction}. Obviously such
proposition has chance of 50{\%}, but on the level of subjective
probabilities, this corresponds not to Hugh, but to his laboratory
assistant.

\section{Discussion}

The difference between internal and external viewpoints regarding
the relationship between observers and the quantum description
(cf.\ Tegmark 1998) becomes starkly clear when compared to a
similar apparent quantum-mechanical PP violation, now in the
context of the "collapse" theories.

Bostrom (2001) describes a "Quantum Joe" thought experiment in
which it is explicitly stipulated that "no-collapse" theories are
false, and that the wavefunction collapse is real and stochastic
(as in the orthodox Copenhagen form of quantum mechanics):

\begin{quotation}
\noindent \textit{Quantum Joe}

\noindent Joe, the amateur scientist, has discovered that he is
alone in the cosmos so far. He builds a quantum device which
according to quantum physics has a one-in-ten chance of outputting
any single-digit integer. He also builds a reproduction device
which when activated will create ten thousand clones of Joe. He
then hooks up the two so that the reproductive device will kick
into action unless the quantum device outputs a zero; but if the
output is a zero, then the reproductive machine will be destroyed.
There are not enough materials left for Joe to reproduce in some
other way, so he will then have been the only observer... Using
the same kinds of argument as before, we can show that Joe should
expect that a zero come up, even though the objective (physical)
chance is a mere 10{\%}.
\end{quotation}

Here we also perceive the apparent violation of PP, due to
probabilistic expectations following the so-called {\it
Self-Sampling Assumption\/} (i.e.\ the assumption that every
observer should reason as if they were a random sample drawn from
the set of all observers). The violation is only apparent, since
it is based on indexical information about Joe's position in the
human species, which is deemed inadmissible. In contradistinction
to the Quantum Joe thought experiment, in the Quantum Hugh case
the truth of "no-collapse" quantum mechanics is explicitly
stipulated. Note that if some of the ``collapse'' theories, say
the orthodox one or the one of Ghirardi, Rimini and Weber (1986),
is correct description of reality, the validity of PP is
straightforward and obvious. On the other hand, both sorts of
apparent conflict, the one in the quantum suicide and the one in
the Quantum Joe experiment, are not at all obvious, and ultimately
depend on considerations related to the properties of intelligent
observers themselves. It is interesting to speculate whether there
are other sorts of "anthropic" evidence in different physical
setups, but this goes beyond the scope of the present study.

We conclude that even in the case of Everett's (or any other
"no-collapse") theory being the correct description of reality, we
can---with our experimenter---coherently believe in equating
subjective credence with objective chance in quantum-mechanical
experiments. This is true provided our \textit{insight\/} in the
nature of experiment, definition of relevant concepts in the
multiverse settings of the "no-collapse" theories, and relevant
capacities of the experimenter is deep enough. We can then, in
general, retain both PP and the "no-collapse" theories.

\vspace{0.5cm}

\noindent \textbf{Acknowledgements.} Two anonymous referees
contributed interesting and provoking comments and criticisms,
which contributed to the quality of the present manuscript.

\vspace{1cm}

\section*{References}

\refe Bostrom, N. 2001, \textit{Synthese\/} \textbf{127}, 359.

\refe Bostrom, N. 2002, \textit{Anthropic Bias: Observation
Selection Effects\/} (Routledge, New York).

\refe Deutsch, D. 1985, {\it International Journal of the
Theoretical Physics\/} {\bf 24}, 1.

\refe Everett, H. III 1957, \textit{Reviews of Modern Physics\/}
\textbf{29}, 454.

\refe Ghirardi, G. C., Rimini, A. and Weber, T. 1986,
\textit{Physical Review D\/} \textbf{34}, 470.

\refe Lewis, D. 1986, \textit{Philosophical Papers\/} (Oxford
University Press, New York).

\refe Lewis, D. 2004, \textit{Australasian Journals of
Philosophy\/} {\bf 82}, 3-22.

\refe Lewis, P. J. 2000, \textit{Analysis\/} {\bf 60}, 22-29.

\refe Moravec, H. P. 1988, \textit{Mind Children: The Future of
Robot and Human Intelligence\/} (Harvard University Press,
Cambridge).

\refe Papineau, D. 2003, \textit{Analysis\/} {\bf 63}, 51-58.

\refe Papineau, D. 2004, \textit{Australasian Journals of
Philosophy\/} {\bf 82}, 153-169.

\refe Parfit, D. 1984, \textit{Reasons and Persons\/} (Oxford
University Press, Oxford).

\refe Price, H. 1996, \textit{Time's Arrow and Archimedes'
Point\/} (Oxford University Press, Oxford).

\refe Squires, E. 1986, \textit{The Mystery of the Quantum
World\/} (Institute of Physics Publishing, Bristol).

\refe Tappenden, P. 2000, \textit{British Journal for the
Philosophy of Science\/} \textbf{51}, 99.

\refe Tappenden, P. 2004, \textit{Analysis\/} {\bf 64}, 157-164.

\refe Tegmark, M. 1998, \textit{Fortschritte der Physik\/}
\textbf{46}, 855.

\refe Vranas, P. 1998, in \textit{Sixteenth Biennial Meeting of
the Philosophy of Science Association}, Kansas City, Missouri.

\refe Zeh, D. 1992, \textit{The Physical Basis of the Direction of
Time\/} (Springer-Verlag, Berlin).

\end{document}